\documentclass[letterpaper, 12pt]{article}[2000/05/19]
\usepackage[english]{babel}
\usepackage{amsfonts,amsmath,amssymb,amsthm,latexsym,amscd,mathrsfs}
\usepackage{ifthen,cite}
\usepackage[bookmarksnumbered=true]{hyperref}

\hypersetup{pdfpagetransition={Split}}

\newcommand{\evenhead}{Author \ name}
\newcommand{\oddhead}{Article \ name}
\newcommand{\theArticleName}{Article \ name}

\newcommand{\FirstPageHeading}[1]{\thispagestyle{empty}%
\noindent\raisebox{0pt}[0pt][0pt]{\makebox[\textwidth]{\protect\footnotesize \sf }}\par}

\newcommand{\ArticleName}[1]{\renewcommand{\theArticleName}{#1}\vspace{-2mm}\par\noindent {\LARGE\bf  #1\par}}
\newcommand{\Author}[1]{\vspace{5mm}\par\noindent {\Large  #1\par} \par\vspace{2mm}\par}
\newcommand{\Address}[1]{\vspace{2mm}\par\noindent {\it #1} \par}
\newcommand{\Email}[1]{\ifthenelse{\equal{#1}{}}{}{\par\noindent {\rm E-mail: }{\it  #1} \par}}
\newcommand{\URLaddress}[1]{\ifthenelse{\equal{#1}{}}{}{\par\noindent {\rm URL: }{\tt  #1} \par}}
\newcommand{\EmailD}[1]{\ifthenelse{\equal{#1}{}}{}{\par\noindent {$\phantom{\dag}$~\rm E-mail: }{\it  #1} \par}}
\newcommand{\URLaddressD}[1]{\ifthenelse{\equal{#1}{}}{}{\par\noindent {$\phantom{\dag}$~\rm URL: }{\tt  #1} \par}}

\newcommand{\Abstract}[1]{\vspace{6mm}\par\noindent\hspace*{10mm}
\parbox{140mm}{\small {\bf Abstract.} #1}\par}
\newcommand{\Keywords}[1]{\vspace{3mm}\par\noindent\hspace*{10mm}
\parbox{140mm}{\small {\bf Key words:} \rm #1}\par}
\newcommand{\Classification}[1]{\vspace{3mm}\par\noindent\hspace*{10mm}
\parbox{140mm}{\small {\it 2000 Mathematics Subject Classification:} \rm #1}\vspace{3mm}\par}
\newcommand{\ShortArticleName}[1]{\renewcommand{\oddhead}{#1}}
\newcommand{\AuthorNameForHeading}[1]{\renewcommand{\evenhead}{#1}}

\long\def\@makecaption#1#2{
  \sbox\@tempboxa{\small \textbf{#1.}\ \ #2}%
  \ifdim \wd\@tempboxa >\hsize
    {\small \textbf{#1.}\ \ #2}\par \else
    \global \@minipagefalse
    \hb@xt@\hsize{\hfil\box\@tempboxa\hfil}%
  \fi \vskip\belowcaptionskip}

\def\numberwithin#1#2{\@ifundefined{c@#1}{\@nocounterr{#1}}{%
  \@ifundefined{c@#2}{\@nocnterr{#2}}{%
  \@addtoreset{#1}{#2}%
  \toks@\@xp\@xp\@xp{\csname the#1\endcsname}%
  \@xp\xdef\csname the#1\endcsname
    {\@xp\@nx\csname the#2\endcsname.\the\toks@}}}}
\def\E^#1{{\buildrel #1 \over\vee}}

{\theoremstyle{definition}

}

\usepackage{tikz}

\begin{document}

\FirstPageHeading{V.I. Gerasimenko}

\AuthorNameForHeading{V.I. Gerasimenko}

\ShortArticleName{Kinetic equations of active soft matter}

\ArticleName{\textcolor{blue!50!black}{On Kinetic Equations for Collisional Dynamics \\of Active Soft Condensed Matter}}

\Author{V.I. Gerasimenko$^\ast$\footnote{E-mail: \emph{gerasym@imath.kiev.ua}}}

\Address{$^\ast$Institute of Mathematics of the NAS of Ukraine,\\
   3, Tereshchenkivs'ka Str., 01601, Kyiv-4, \\ Ukraine}

\bigskip

\Abstract{We consider a new approach to the description of the collective behavior
of complex systems of mathematical biology based on the evolution equations for
observables of such systems. This representation of the kinetic evolution seems, in fact,
the direct mathematically fully consistent formulation modeling the collective behavior
of biological systems since the traditional notion of the state in kinetic theory is more
subtle and it is an implicit characteristic of the populations of living creatures.
}

\bigskip

\Keywords{active soft matter, marginal observable, correlation function, kinetic equation, scaling limit.}
\vspace{2pc}
\Classification{35Q92; 37N25; 82C22.}

\makeatletter
\renewcommand{\@evenhead}{
\hspace*{-3pt}\raisebox{-7pt}[\headheight][0pt]{\vbox{\hbox to \textwidth {\thepage \hfil \evenhead}\vskip4pt \hrule}}}
\renewcommand{\@oddhead}{
\hspace*{-3pt}\raisebox{-7pt}[\headheight][0pt]{\vbox{\hbox to \textwidth {\oddhead \hfil \thepage}\vskip4pt\hrule}}}
\renewcommand{\@evenfoot}{}
\renewcommand{\@oddfoot}{}
\makeatother

\newpage
\vphantom{math}

\protect\textcolor{blue!50!black}{\tableofcontents}
\vspace{0.5cm}

\textcolor{blue!50!black}{\section{Introduction}}

The rigorous derivation of kinetic equations for soft condensed matter remains an open
problem so far. It should be noted wide applications of these evolution equations to the
description of collective processes of various nature \cite{SR12}-\cite{ALBA}, in particular,
the collective behavior of complex systems of mathematical biology \cite{L08}-\cite{CDW}.
We emphasize that the considerable advance in solving the problem of rigorous modeling
of the kinetic evolution of systems with a large number of constituents (entities) of
mathematical biology, in particular, systems of large number of cells, is recently
observed \cite{MCV}-\cite{GF13} (and see references cited therein).

In modern research, the main approach to the problem of the rigorous derivation of kinetic
equation consists in the construction of scaling limits of a solution of evolution
equations which describe the evolution of states of a many-particle system, in particular,
a perturbative solution of the corresponding BBGKY hierarchy \cite{CGP97}-\cite{CIP}.

In this paper, we review a new approach to the description of the collective behavior
of complex systems of mathematical biology \cite{M15},\cite{VZ} within the framework
of the evolution of observables. This representation of the kinetic evolution seems, in
fact, the direct mathematically fully consistent formulation modeling kinetic evolution
of biological systems since the notion of the state is more subtle and it is an implicit
characteristic of populations of living creatures.

One of the advantages of the developed approach is the opportunity to construct kinetic
equations in scaling limits, involving initial correlations, in particular, that can
characterize the condensed states of soft matter. We note also that such approach
is also related to the problem of a rigorous derivation of the non-Markovian kinetic-type
equations from underlying many-cell dynamics which make it possible to describe the memory
effects of the kinetic evolution of cells.

Using suggested approach, we establish a mean field asymptotic behavior of the hierarchy
of evolution equations for marginal observables of a large system of interacting stochastic
processes of collisional kinetic theory \cite{L11}, modeling the microscopic evolution of
active soft condenced matter \cite{ALBA},\cite{MJRLPRAS}. The constructed scaling limit of
a nonperturbative solution of this hierarchy is governed by the set of recurrence evolution
equations, namely, by the dual Vlasov hierarchy for interacting stochastic processes.

Furthermore, we established that for initial states specified by means of a one-particle
distribution function and correlation functions the evolution of additive-type marginal
observables is equivalent to a solution of the Vlasov-type kinetic equation with initial
correlations, and a mean field asymptotic behavior of non-additive type marginal observables
is equivalent to the sequence of explicitly defined correlation functions which describe the
propagation of initial correlations of active soft condensed matter.


\textcolor{blue!50!black}{\section{On collisional dynamics of active soft condensed matter
and the evolution of marginal observables}}
The many-constituent systems of active soft condensed matter \cite{ALBA},\cite{MJRLPRAS} are dynamical
systems displaying a collective behavior which differs from the statistical behavior of usual
gases\cite{CGP97},\cite{CIP}. In the first place, their own distinctive features are connected
with the fact that their constituents (entities or self-propelled particles) show the ability
to retain various complexity features \cite{ALBA}-\cite{VZ}. To specify such nature of entities
we consider the dynamical system suggested in papers \cite{L08},\cite{L11},\cite{LP} which is
based on the Markov jump processes that must represent the intrinsic properties of living creatures.

A description of many-constituent systems is formulated in terms of two sets of objects: observables
and states. The functional of the mean value of observables defines a duality between observables
and states and as a consequence there exist two approaches to the description of the evolution
of such systems, namely in terms of the evolution equations for observables and for states. In
this section, we adduce some preliminary facts about dynamics of finitely many entities of various
subpopulations described within the framework of nonequilibrium grand canonical ensemble \cite{CGP97}.

We consider a system of entities of various $M$ subpopulations introduced in paper \cite{L11}
in case of non-fixed, i.e. arbitrary, but finite average number of entities. Every $i$th
entity is characterized by: $\textbf{u}_i=(j_i,u_i)\in\mathcal{J}\times\mathcal{U}$, where
$j_i\in\mathcal{J}\equiv(1,\ldots,M)$ is a number of its subpopulation, and
$u_i\in\mathcal{U}\subset\mathbb{R}^{d}$ is its microscopic state \cite{L11}. The stochastic
dynamics of entities of various subpopulations is described by the semigroup
$e^{t\Lambda}=\oplus_{n=0}^\infty e^{t\Lambda_n}$ of the Markov jump process defined on the
space $C_\gamma$ of sequences  $b=(b_0,b_1,\ldots,b_n,\ldots)$ of measurable bounded
functions $b_n(\textbf{u}_1,\ldots,\textbf{u}_n)$ that are symmetric with respect to
permutations of the arguments $\textbf{u}_1,\ldots,\textbf{u}_n$ and equipped with the norm:
\begin{eqnarray*}
    &&\|b\|_{C_\gamma}=\max_{n\geq0}\frac{\gamma^n}{n!}\|b_n\|_{C_n}=
      \max_{n\geq0}\frac{\gamma^n}{n!}\max_{j_1,\ldots,j_n}\max_{u_1,\ldots,u_n}
      \big|b_n(\textbf{u}_1,\ldots,\textbf{u}_n)\big|,
\end{eqnarray*}
where $\gamma<1$ is a parameter. The infinitesimal generator $\Lambda_n$ of collisional dynamics
(the Liouville operator of $n$ entities) is defined on the subspace $C_n$ of the space
$C_\gamma$ and it has the following structure \cite{L11}:
\begin{eqnarray}\label{gen_obs_gener}
    &&\hskip-5mm(\Lambda_n b_n)(\textbf{u}_1,\ldots,\textbf{u}_n)\doteq
        \sum_{m=1}^M \varepsilon^{m-1}\sum_{i_1\neq\ldots\neq i_m=1}^n
        (\Lambda^{[m]}(i_1,\ldots,i_m)b_n)(\textbf{u}_1,\ldots,\textbf{u}_n)=\\
    &&\hskip-5mm=\sum_{m=1}^M \varepsilon^{m-1}\sum_{i_1\neq\ldots\neq i_m=1}^n
        a^{[m]}(\textbf{u}_{i_1},\ldots,\textbf{u}_{i_m})\big(\int_{\mathcal{J}\times\mathcal{U}}
        A^{[m]}(\textbf{v};\textbf{u}_{i_1},\ldots,\textbf{u}_{i_m})\times\nonumber\\
    &&\hskip+5mm\times b_n(\textbf{u}_1,\ldots,\textbf{u}_{i_1-1},\textbf{v},\textbf{u}_{i_1+1},
        \ldots\textbf{u}_n)d\textbf{v}-b_n(\textbf{u}_1,\ldots,\textbf{u}_n)\big)\nonumber,
\end{eqnarray}
where $\varepsilon>0$ is a scaling parameter \cite{BL14}, the functions
$a^{[m]}(\textbf{u}_{i_1},\ldots,\textbf{u}_{i_m}),\,m\geq1,$ characterize the interaction between
entities, in particular, in case of $m=1$ it is the interaction of entities with an external environment.
These functions are measurable positive bounded functions on $(\mathcal{J}\times\mathcal{U})^n$ such that:
\begin{eqnarray*}
   &&0\leq a^{[m]}(\textbf{u}_{i_1},\ldots,\textbf{u}_{i_m})\leq a^{[m]}_*,
\end{eqnarray*}
where $a^{[m]}_*$ is some constant. The functions
$A^{[m]}(\textbf{v};\textbf{u}_{i_1},\ldots,\textbf{u}_{i_m}),\,m\geq1$, are measurable positive
integrable functions which describe the probability of the transition of the $i_1$ entity in the
microscopic state $u_{i_1}$ to the state $v$ as a result of the interaction with entities in the
states $u_{i_2},\ldots,u_{i_m}$ (in case of $m=1$ it is the interaction with an external environment).
The functions $A^{[m]}(\textbf{v};\textbf{u}_{i_1},\ldots,\textbf{u}_{i_m}),\,m\geq1$, satisfy
the conditions:
\begin{eqnarray*}
   &&\int_{\mathcal{J}\times\mathcal{U}}A^{[m]}(\textbf{v};\textbf{u}_{i_1},\ldots,
      \textbf{u}_{i_m})d\textbf{v}=1, \quad m\geq1.
\end{eqnarray*}
We refer to paper \cite{L11}, where examples of the functions $a^{[m]}$ and $A^{[m]}$ are given
in the context of biological systems.

In case of $M=1$ generator (\ref{gen_obs_gener}) has the form $\sum_{i_1=1}^n\Lambda^{[1]}_n(i_1)$
and it describes the free stochastic evolution of entities, i.e. the evolution of self-propelled particles.
The case of $M=m\geq2$ corresponds to a system with the $m$-body interaction of entities in the sense
accepted in kinetic theory \cite{G12}.
The $m$-body interaction of entities is the distinctive property of biological systems in comparison
with many-particle systems, for example, gases of atoms with a pair interaction potential.

On the space $C_n$ the one-parameter mapping $e^{t\Lambda_n}$ is a bounded $\ast$-weak continuous
semigroup of operators.

The observables of a system of a non-fixed number of entities of various subpopulations are the
sequences $O=(O_{0},O_{1}(\textbf{u}_1),\ldots,O_{n}(\textbf{u}_1,\ldots,\textbf{u}_n),\ldots)$ of
functions $O_{n}(\textbf{u}_1,\ldots,\textbf{u}_n)$ defined on $(\mathcal{J}\times\mathcal{U})^n$
and $O_{0}$ is a real number. The evolution of observables is described by the sequences
$O(t)=(O_{0},O_{1}(t,\textbf{u}_1),\ldots,O_{n}(t,\textbf{u}_1,\ldots,\textbf{u}_n),\ldots)$
of the functions
\begin{eqnarray*}
   &&O_{n}(t)=e^{t\Lambda_n}O_{n}^0, \quad n\geq1,
\end{eqnarray*}
i.e. they are the corresponding solution of the Cauchy problem of the Liouville equations
(or the Kolmogorov forward equation) with corresponding initial data $O_{n}^0$:
\begin{eqnarray*}
   &&\frac{\partial}{\partial t}O_{n}(t)=\Lambda_nO_{n}(t),\\
   &&O_{n}(t)\mid_{t=0}=O_{n}^0, \quad n\geq1,
\end{eqnarray*}
or in case of $n$ noninteracting entities (self-propelled particles) these equations have the form
\begin{eqnarray*}
    &&\hskip-7mm\frac{\partial}{\partial t}O_{n}(t,\textbf{u}_1,\ldots,\textbf{u}_n)=\sum_{i=1}^n
        a^{[1]}(\textbf{u}_{i})\big(\int_{\mathcal{J}\times\mathcal{U}}
        A^{[1]}(\textbf{v};\textbf{u}_{i})O_{n}(t,\textbf{u}_1,\ldots,\textbf{u}_{i-1},\textbf{v},\textbf{u}_{i+1},
        \ldots\textbf{u}_n)d\textbf{v}-\\
    &&\hskip+7mm-O_{n}(t,\textbf{u}_1,\ldots,\textbf{u}_n)\big), \quad n\geq1.
\end{eqnarray*}

The average values of observables (mean values of observables) are determined by the following
positive continuous linear functional defined on the space $C_\gamma$:
\begin{eqnarray}\label{averageD}
     &&\hskip-7mm\langle O\rangle(t)=(I,D(0))^{-1}(O(t),D(0))\doteq
        (I,D(0))^{-1}\sum\limits_{n=0}^{\infty}\frac{1}{n!}
        \int_{(\mathcal{J}\times\mathcal{U})^n} d\textbf{u}_1\ldots d\textbf{u}_{n}\,O_{n}(t)\,D_{n}^0,
\end{eqnarray}
where $D(0)=(1,D_{1}^0,\ldots,D_{n}^0,\ldots)$ is a sequence of nonnegative functions
$D_{n}^0$ defined on $(\mathcal{J}\times\mathcal{U})^n$ that describes the states of
a system of a non-fixed number of entities of various subpopulations at initial time
and $(I,D(0))=\sum_{n=0}^{\infty}\frac{1}{n!}\int_{(\mathcal{J}\times\mathcal{U})^n}d\textbf{u}_1\ldots
d\textbf{u}_{n}\,D_{n}^0$ is a normalizing factor (the grand canonical partition function).

Let $L^{1}_{\alpha}=\oplus^{\infty}_{n=0}\alpha^n L^{1}_{n}$ be the space of sequences
$f=(f_0,f_1,\ldots,f_n,\ldots)$ of the integrable functions $f_n(\textbf{u}_1,\ldots,\textbf{u}_n)$
defined on $(\mathcal{J}\times\mathcal{U})^n$, that are symmetric with respect to permutations
of the arguments $\textbf{u}_1,\ldots,\textbf{u}_n$, and equipped with the norm:
\begin{eqnarray*}
   &&\|f\|_{L^{1}_\alpha}=\sum\limits_{n=0}^\infty\alpha^n\|f_n\|_{L^{1}_n}=
       \sum\limits_{n=0}^\infty\alpha^n\sum\limits_{j_1\in\mathcal{J}}\ldots
       \sum\limits_{j_n\in\mathcal{J}}\,\,
       \int_{\mathcal{U}^n}du_1\ldots du_{n}\big|f_n(\textbf{u}_1,\ldots,\textbf{u}_n)\big|,
\end{eqnarray*}
where $\alpha>1$ is a parameter. Then for $D(0)\in L^{1}$ and $O(t)\in C_\gamma$ average value
functional (\ref{averageD}) exists and it determines a duality between observables and states.

As a consequence of the validity for functional (\ref{averageD}) of the following equality:
\begin{eqnarray*}
     &&(I,D(0))^{-1}(O(t),D(0))=(I,D(0))^{-1}(e^{t\Lambda}O(0)\,D(0))=\\
     &&=(I,e^{t\Lambda^\ast}D(0))^{-1}(O(0)\,e^{t\Lambda^\ast}D(0))\equiv(I,D(t))^{-1}(O(0),D(t)),
\end{eqnarray*}
where $e^{t\Lambda^\ast}=\oplus^{\infty}_{n=0}e^{t\Lambda_n^\ast}$ is the adjoint semigroup
of operators with respect to the semigroup $e^{t\Lambda}=\oplus^{\infty}_{n=0}e^{t\Lambda_n}$,
it is possible to describe the evolution within the framework of the evolution of states.
Indeed, the evolution of all possible states, i.e. the sequence
$D(t)=(1,D_{1}(t,\textbf{u}_1),\ldots, D_{n}(t,\textbf{u}_1,\ldots,$ $\textbf{u}_n),\ldots)\in L^{1}$
of the distribution functions $D_{n}(t),\, n\geq1$, is determined by the formula:
\begin{eqnarray*}
   &&D_{n}(t)=e^{t\Lambda^\ast_n}D^0_n, \quad n\geq1,
\end{eqnarray*}
where the generator $\Lambda^\ast_n$ is the adjoint operator to operator
(\ref{gen_obs_gener}) and on $L^{1}_{n}$ it is defined as follows:
\begin{eqnarray}\label{gen_state_gener}
 &&\hskip-7mm (\Lambda^\ast_n f_n)(\textbf{u}_1,\ldots,\textbf{u}_n)\doteq
     \sum_{m=1}^M\varepsilon^{m-1}\sum_{i_1\neq\ldots\neq i_m=1}^n
     \big(\int_{\mathcal{J}\times\mathcal{U}}
     A^{[m]}(\textbf{u}_{i_1};\textbf{v},\textbf{u}_{i_2},\ldots,\textbf{u}_{i_m})a^{[m]}(\textbf{v},\\
 &&\hskip-7mm\textbf{u}_{i_2},\ldots,\textbf{u}_{i_m})f_n(\textbf{u}_1,\ldots,
     \textbf{u}_{{i_1}-1},\textbf{v},\textbf{u}_{{i_1}+1},\ldots,\textbf{u}_n)d\textbf{v}-
     a^{[m]}(\textbf{u}_{i_1},\ldots,\textbf{u}_{i_m})f_n(\textbf{u}_1,\ldots,\textbf{u}_n)\big)\nonumber,
\end{eqnarray}
where the functions $A^{[m]}$ and $a^{[m]}$ are defined as above in (\ref{gen_obs_gener}).

The function $D_{n}(t)$ is a solution of the Cauchy problem of the dual Liouville equation
(or the Kolmogorov backward equation).

On the space $L^{1}_{n}$ the one-parameter mapping $e^{t\Lambda^\ast_n}$ is a bounded strong continuous
semigroup of operators \cite{GF13}.

For the description of microscopic behavior of many-entity systems we also introduce the hierarchies
of evolution equations for marginal observables and marginal distribution functions known as the dual
BBGKY hierarchy and the BBGKY hierarchy, respectively \cite{GF13}. These hierarchies are constructed
as the evolution equations for one more method of the description of observables and states of finitely
many entities.

An equivalent approach to the description of observables and states of many-entity systems
is given in terms of marginal observables
$B(t)=(B_0,B_{1}(t,\textbf{u}_1),\ldots,B_{s}(t,\textbf{u}_1,\ldots,\textbf{u}_s),\ldots)$
and marginal distribution functions
$F(0)=(1,F_{1}^{0,\varepsilon}(\textbf{u}_1),\ldots,F_{s}^{0,\varepsilon}(\textbf{u}_1,\ldots,\textbf{u}_s),\ldots)$.
Considering formula (\ref{averageD}), marginal observables and marginal distribution functions
are introduced according to the equality:
\begin{eqnarray}\label{avmar}
   &&\big\langle O\big\rangle(t)=(I,D(0))^{-1}(O(t),D(0))=(B(t),F(0)),
\end{eqnarray}
where $(I,D(0))$ is a normalizing factor defined as above. If $F(0)\in L^{1}_{\alpha}$ and $B(0)\in C_\gamma$,
then at $t\in \mathbb{R}$ the functional $(B(t),F(0))$ exists under the condition that: $\gamma>\alpha^{-1}$.

Thus, the relationship of marginal distribution functions $F(0)=(1,F_{1}^{0,\varepsilon},\ldots,F_{s}^{0,\varepsilon},\ldots)$
and the distribution functions $D(0)=(1,D_{1}^0,\ldots,D_{n}^0,\ldots)$ is determined by the formula:
\begin{eqnarray*}\label{ms}
      &&\hskip-5mm F_{s}^{0,\varepsilon}(\textbf{u}_1,\ldots,\textbf{u}_s)\doteq
          (I,D(0))^{-1}\sum\limits_{n=0}^{\infty}\frac{1}{n!}\,
          \int_{(\mathcal{J}\times\mathcal{U})^n}d\textbf{u}_{s+1}\ldots
          d\textbf{u}_{s+n}\,D_{s+n}^0(\textbf{u}_1,\ldots,\textbf{u}_{s+n}), \quad s\geq 1,
\end{eqnarray*}
and, respectively, the marginal observables are determined in terms of observables as follows:
\begin{eqnarray*}\label{mo}
      &&\hskip-5mm B_{s}(t,\textbf{u}_1,\ldots,\textbf{u}_s)\doteq\sum_{n=0}^s\,
            \frac{(-1)^n}{n!}\sum_{j_1\neq\ldots\neq j_{n}=1}^s
            O_{s-n}\big(t,(\textbf{u}_1,\ldots,\textbf{u}_s)\setminus(\textbf{u}_{j_1},\ldots,\textbf{u}_{j_{n}})\big),
            \quad s\geq 1.
\end{eqnarray*}

Two equivalent approaches to the description of the evolution of many interacting entities
are the consequence of the validity of the following equality for the functional of mean values
of marginal observables:
\begin{eqnarray*}
   &&(B(t),F(0))=(B(0),F(t)),
\end{eqnarray*}
where $B(0)=(1,B_{1}^{0,\varepsilon},\ldots,B_{s}^{0,\varepsilon},\ldots)$ is a sequence of marginal
observables at initial moment.

We remark that the evolution of many-entity systems is usually described
within the framework of the evolution of states by the sequence
$F(t)=(1,F_{1}(t,\textbf{u}_1),\ldots,F_{s}(t,\textbf{u}_1,\ldots,\textbf{u}_s),\ldots)$
of marginal distribution functions $F_s(t,\textbf{u}_1,\ldots,\textbf{u}_s)$ governed by
the BBGKY hierarchy for interacting entities.

The evolution of a non-fixed number of interacting entities of various subpopulations within the framework
of marginal observables (\ref{mo}) is described by the Cauchy problem of the dual BBGKY hierarchy \cite{BG}:
\begin{eqnarray}\label{adh}
   &&\hskip-5mm \frac{d}{dt}B(t)=\Lambda+\sum\limits_{n=1}^{\infty}\frac{1}{n!}
     \big[\ldots\big[\Lambda,\underbrace{\mathfrak{a}^{+} \big],\ldots,\mathfrak{a}^{+}}_{\hbox{n-times}}\big]B(t),\\
       \nonumber\\
   \label{dhi}
   &&\hskip-5mm B(t)|_{t=0}=B(0),
\end{eqnarray}
where on $C_{\gamma}$ the operator $\mathfrak{a}^{+}$ (an analog of the creation operator) is defined as follows
\begin{eqnarray*}\label{oper_znuw}
   &&(\mathfrak{a}^{+}b)_{s}(\textbf{u}_1,\ldots,\textbf{u}s)\doteq
        \sum_{j=1}^s\,b_{s-1}(\textbf{u}_1,\ldots,\textbf{u}_{j-1},\textbf{u}_{j+1},\ldots,\textbf{u}_s),
\end{eqnarray*}
the operator $\Lambda=\oplus_{n=0}^\infty \Lambda_n$ is defined by (\ref{gen_obs_gener}),
and the symbol $\big[\,\cdot\,,\,\cdot\,\big]$ denotes the commutator of operators.

In the componentwise form, the abstract hierarchy (\ref{adh}) has the form:
\begin{eqnarray}\label{dh}
   &&\hskip-5mm \frac{\partial}{\partial t}B_{s}(t,\textbf{u}_1,\ldots,\textbf{u}_s)=
       \Lambda_s B_{s}(t,\textbf{u}_1,\ldots,\textbf{u}_s)+
       \sum\limits_{n=1}^{s}\frac{1}{n!}\sum\limits_{k=n+1}^s \frac{1}{(k-n)!}\times\\
   &&\hskip-5mm \times\sum_{j_1\neq\ldots\neq j_{k}=1}^s\varepsilon^{k-1}\Lambda^{[k]}(j_1,\ldots,
       j_{k})\sum_{i_1\neq\ldots\neq i_{n}\in(j_1,\ldots,j_{k})}B_{s-n}(t,(\textbf{u}_1,\ldots,
       \textbf{u}_s)\setminus(\textbf{u}_{i_1},\ldots,\textbf{u}_{i_{n}})),\nonumber\\
       \nonumber\\
   \label{dhi}
   &&\hskip-5mm B_s(t,\textbf{u}_1,\ldots,\textbf{u}_s)|_{t=0}=
       B_{s}^{0,\varepsilon}(\textbf{u}_1,\ldots,\textbf{u}_s), \quad s\geq 1,
\end{eqnarray}
where the operators $\Lambda_s$ and $\Lambda^{[k]}$ are defined by formulas (\ref{gen_obs_gener}) and
the functions $B_{s}^{0,\varepsilon},\,s\geq 1,$ are a scaled initial data.

A solution $B(t)=(B_{0},B_{1}(t,\textbf{u}_1),\ldots,B_{s}(t,\textbf{u}_1,\ldots,$ $\textbf{u}_s),\ldots)$
of the Cauchy problem of recurrence evolution equations (\ref{dh}),(\ref{dhi}) is given by the following
expansions \cite{GF13}:
\begin{eqnarray}\label{sdh}
   &&\hskip-5mm B_{s}(t,\textbf{u}_1,\ldots,\textbf{u}_s)=
      \sum_{n=0}^s\,\frac{1}{n!}\sum_{j_1\neq\ldots\neq j_{n}=1}^s\mathfrak{A}_{1+n}(t,\{Y\setminus Z\},\,Z)\,
      B_{s-n}^{0,\varepsilon}(\textbf{u}_1,\ldots,\\
   &&\hskip+5mm\textbf{u}_{j_1-1},\textbf{u}_{j_1+1},\ldots,\textbf{u}_{j_n-1},
      \textbf{u}_{j_n+1},\ldots,\textbf{u}_s), \quad s\geq 1,\nonumber
\end{eqnarray}
where the $(1+n)th$-order cumulant of the semigroups $\{e^{t\Lambda_{k}}\}_{t\in\mathbb{R}},\, k\geq1,$
is determined by the formula \cite{BG}:
\begin{eqnarray}\label{cumulantd}
   &&\hskip-5mm\mathfrak{A}_{1+n}(t,\{Y\setminus Z\},\,Z)\doteq
       \sum\limits_{\mathrm{P}:\,(\{Y\setminus Z\},\,Z)={\bigcup}_i Z_i}
       (-1)^{\mathrm{|P|}-1}({\mathrm{|P|}-1})!\prod_{Z_i\subset \mathrm{P}}e^{t\Lambda_{|\theta(Z_i)|}},
\end{eqnarray}
the sets of indexes are denoted by $Y\equiv(1,\ldots,s)$, $Z\equiv(j_1,\ldots,j_{n})\subset Y$, the set
$\{Y\setminus Z\}$ consists from one element $Y\setminus Z=(1,\ldots,j_1-1,j_1+1,\ldots,j_n-1,j_n+1,\ldots,s)$
and the mapping $\theta(\cdot)$ is the declusterization operator defined as follows:
$\theta(\{Y\setminus Z\},\,Z)=Y$.

The simplest examples of expansions for marginal observables (\ref{sdh}) have the following form:
\begin{eqnarray*}
   &&B_{1}(t,\textbf{u}_1)=\mathfrak{A}_{1}(t,1)B_{1}^{\epsilon,0}(\textbf{u}_1),\\
   &&B_{2}(t,\textbf{u}_1,\textbf{u}_2)=\mathfrak{A}_{1}(t,\{1,2\})B_{2}^{\epsilon,0}(\textbf{u}_1,\textbf{u}_2)+
      \mathfrak{A}_{2}(t,1,2)(B_{1}^{\epsilon,0}(\textbf{u}_1)+B_{1}^{\epsilon,0}(\textbf{u}_2)),
\end{eqnarray*}
and, respectively:
\begin{eqnarray*}
   &&\mathfrak{A}_{1}(t,\{1,2\})=e^{t\Lambda_2(1,2)},\\
   &&\mathfrak{A}_{2}(t,1,2)=e^{t\Lambda_2(1,2)}-e^{t\Lambda_1(1)}e^{t\Lambda_1(2)} .
\end{eqnarray*}

For initial data $B(0)=(B_{0},B_{1}^{0,\varepsilon},\ldots,B_{s}^{0,\varepsilon},\ldots)\in C_{\gamma}$
the sequence $B(t)$ of marginal observables given by expansions (\ref{sdh}) is a classical solution
of the Cauchy problem of the dual BBGKY hierarchy for interacting entities (\ref{dh}),(\ref{dhi}).

We note that a one-component sequence of marginal observables corresponds to observables of certain
structure, namely the marginal observable $B^{(1)}(0)=(0,b_{1}^{\epsilon}(\textbf{u}_1),0,\ldots)$ corresponds
to the additive-type observable, and a one-component sequence of marginal observables
$B^{(k)}(0)=(0,\ldots,0,b_{k}^{\epsilon}(\textbf{u}_1,\ldots,\textbf{u}_k),0,\ldots)$ corresponds to the $k$-ary-type
observable \cite{BG}. If in capacity of initial data (\ref{dhi}) we consider the additive-type marginal
observables, then the structure of solution expansion (\ref{sdh}) is simplified and attains the form
\begin{eqnarray}\label{af}
     &&B_{s}^{(1)}(t,\textbf{u}_1,\ldots,\textbf{u}_s)=\mathfrak{A}_{s}(t,1,\ldots,s)
              \sum_{j=1}^s b_{1}^{\epsilon}(\textbf{u}_j), \quad s\geq 1.
\end{eqnarray}
In the case of $k$-ary-type marginal observables solution expansion (\ref{sdh}) has the form
\begin{eqnarray}\label{af-k}
     &&\hskip-5mm B_{s}^{(k)}(t,\textbf{u}_1,\ldots,\textbf{u}_s)=\frac{1}{(s-k)!}\sum_{j_1\neq\ldots\neq j_{s-k}=1}^s
      \mathfrak{A}_{1+s-k}\big(t,\{(1,\ldots,s)\setminus (j_1,\ldots,j_{s-k})\},\\
     &&\hskip+12mm j_1,\ldots,j_{s-k}\big)\,
      b_{k}^{\epsilon}(\textbf{u}_1,\ldots,\textbf{u}_{j_1-1},\textbf{u}_{j_1+1},\ldots,
      \textbf{u}_{j_s-k-1},\textbf{u}_{j_s-k+1},\ldots,\textbf{u}_s),\quad s\geq k,\nonumber
\end{eqnarray}
and, if $1\leq s<k$, we have $B_{s}^{(k)}(t)=0$.

We remark also that expansion (\ref{sdh}) can be also represented in the form of the perturbation
(iteration) series \cite{BG} as a result of applying of analogs of the Duhamel equation
to cumulants of semigroups of operators (\ref{cumulantd}).

\textcolor{blue!50!black}{\section{A mean field asymptotic behavior of the marginal observables
and the kinetic evolution of states}}
To consider mesoscopic properties of a large system of interacting entities we develop an
approach to the description of the kinetic evolution within the framework of the evolution
equations for marginal observables. For this purpose we construct the mean field asymptotics
of a solution of the Cauchy problem of the dual BBGKY hierarchy for interacting entities, modeling
of many-constituent systems of active soft condensed matter \cite{GF13},\cite{GF15}.

We restrict ourself by the case of $M=2$ subpopulations to simplify the cumbersome formulas
and consider the mean field scaling limit of nonperturbative solution (\ref{sdh}) of the Cauchy
problem of the dual BBGKY hierarchy for interacting entities (\ref{dh}),(\ref{dhi}).

Let for initial data $B_{s}^{0,\varepsilon}\in C_s$ there exists the limit function
$b_{s}^0\in C_s$
\begin{eqnarray*}\label{asumdin}
    &&\mathrm{w^{\ast}-}\lim\limits_{\varepsilon\rightarrow 0}\big(\varepsilon^{-s}
         B_{s}^{0,\varepsilon}-b_{s}^0\big)=0,\quad s\geq1,
\end{eqnarray*}
then for arbitrary finite time interval there exists a mean field limit of solution (\ref{sdh})
of the Cauchy problem of the dual BBGKY hierarchy for interacting entities  (\ref{dh}),(\ref{dhi})
in the sense of the $\ast$-weak convergence of the space $C_s$
\begin{eqnarray*}\label{asymt}
   && \mathrm{w^{\ast}-}\lim\limits_{\varepsilon\rightarrow 0} \big(\varepsilon^{-s}B_{s}(t)-b_{s}(t)\big)=0,
   \quad s\geq1,
\end{eqnarray*}
where the limit sequence $b(t)=(b_0,b_1(t),\ldots,b_{s}(t),\ldots)$ of  marginal observables
is determined by the following expansions:
\begin{eqnarray}\label{Iterd}
   &&\hskip-9mm b_{s}(t,\textbf{u}_1,\ldots,\textbf{u}_s)=\\
   &&\hskip-7mm=\sum\limits_{n=0}^{s-1}\,\int_0^tdt_{1}\ldots\int_0^{t_{n-1}}dt_{n}
      \,e^{(t-t_{1})\sum\limits_{k_{1}=1}^{s}\Lambda^{[1]}(k_{1})}\sum\limits_{i_{1}\neq j_{1}=1}^{s}
      \Lambda^{[2]}(i_{1},j_{1})e^{(t_{1}-t_{2})\sum\limits_{l_{1}=1,l_{1}\neq j_{1}}^{s}\Lambda^{[1]}(l_{1})}
       \ldots\nonumber\\
   &&\hskip-7mm e^{(t_{n-1}-t_{n})\hskip-1mm\sum\limits^{s}_{\mbox{\scriptsize $\begin{array}{c}k_{n}=1,\\
      k_{n}\neq (j_{1},\ldots,j_{n-1}))\end{array}$}}\hskip-1mm\Lambda^{[1]}(k_{n})}
            \hskip-2mm\sum\limits^{s}_{\mbox{\scriptsize $\begin{array}{c}i_{n}\neq j_{n}=1,\\
      i_{n},j_{n}\neq (j_{1},\ldots,j_{n-1})\end{array}$}}\hskip-1mm\Lambda^{[2]}(i_{n},j_{n})
      e^{t_{n}\hskip-1mm\sum\limits^{s}_{\mbox{\scriptsize $\begin{array}{c}l_{n}=1,\\
      l_{n}\neq (j_{1},\ldots,j_{n}))\end{array}$}}\hskip-1mm\Lambda^{[1]}(l_{n})}
      b_{s-n}^0((\textbf{u}_1,\nonumber\\
   &&\hskip-7mm\ldots,\textbf{u}_s)\setminus (\textbf{u}_{j_{1}},\ldots,\textbf{u}_{j_{n}})),\quad s\geq1.\nonumber
\end{eqnarray}
In particular, the limit marginal observable $b_{s}^{(1)}(t)$ of the additive-type marginal observable
(\ref{af}) is determined as a special case of expansions (\ref{Iterd}):
\begin{eqnarray*}\label{itvad}
   &&\hskip-9mm b_{s}^{(1)}(t,\textbf{u}_1,\ldots,\textbf{u}_s)=\\
   &&\hskip-7mm \int_0^t dt_{1}\ldots\int_0^{t_{s-2}}dt_{s-1}\,
       \,e^{(t-t_{1})\sum\limits_{k_{1}=1}^{s}\Lambda^{[1]}(k_{1})}\sum\limits_{i_{1}\neq j_{1}=1}^{s}
      \Lambda^{[2]}(i_{1},j_{1})e^{(t_{1}-t_{2})\sum\limits_{l_{1}=1,l_{1}\neq j_{1}}^{s}\Lambda^{[1]}(l_{1})}
       \ldots\nonumber\\
   &&\hskip-7mm e^{(t_{s-2}-t_{s-1})\hskip-2mm\sum\limits^{s}_{\mbox{\scriptsize $\begin{array}{c}k_{s-1}=1,\\
      k_{n}\neq (j_{1},\ldots,j_{n-1}))\end{array}$}}\hskip-12mm\Lambda^{[1]}(k_{n})}
            \hskip-2mm\sum\limits^{s}_{\mbox{\scriptsize $\begin{array}{c}i_{s-1}\neq j_{s-1}=1,\\
      i_{s-1},j_{s-1}\neq (j_{1},\ldots,j_{s-2})\end{array}$}}\hskip-2mm\Lambda^{[2]}(i_{s-1},j_{s-1})\,
      e^{t_{s-1}\hskip-2mm\sum\limits^{s}_{\mbox{\scriptsize $\begin{array}{c}l_{s-1}=1,\\
      l_{s-1}\neq (j_{1},\ldots,j_{s-1}))\end{array}$}}\hskip-2mm\Lambda^{[1]}(l_{s-1})}\times\nonumber\\
   &&\hskip-7mm\times b_{1}^{0}\big((\textbf{u}_1,\ldots,\textbf{u}_s)\setminus(\textbf{u}_{j_{1}},\ldots,\textbf{u}_{j_{s-1}})\big),
      \quad s\geq1,\nonumber
\end{eqnarray*}
for example,
\begin{eqnarray*}
   &&\hskip-8mm b_{1}^{(1)}(t,\textbf{u}_1)=e^{t\Lambda^{[1]}(1)}\,b_{1}^{0}(\textbf{u}_1),\\
   &&\hskip-8mm b_{2}^{(1)}(t,\textbf{u}_1,\textbf{u}_2)=\int_0^t dt_{1}\prod\limits_{i=1}^{2}e^{(t-t_{1})\Lambda^{[1]}(i)}\,
      \Lambda^{[2]}(1,2)\sum\limits_{j=1}^{2}e^{t_{1}\Lambda^{[1]}(j)}\,b_{1}^{0}(\textbf{u}_j).
\end{eqnarray*}

The proof of this statement is based on the corresponding formulas for cumulants of asymptotically
perturbed semigroups of operators (\ref{cumulantd}).

If $b^0\in C_{\gamma}$, then the sequence $b(t)=(b_0,b_1(t),\ldots,b_{s}(t),\ldots)$
of limit marginal observables (\ref{Iterd}) is generalized global in time solution of the Cauchy
problem of the dual Vlasov hierarchy:
\begin{eqnarray}\label{vdh}
   &&\hskip-5mm \frac{\partial}{\partial t}b_{s}(t)=
     \sum\limits_{j=1}^{s}\Lambda^{[1]}(j)\,b_{s}(t)+
     \sum_{j_1\neq j_{2}=1}^s\Lambda^{[2]}(j_1,j_{2})\,b_{s-1}(t,\textbf{u}_1,\ldots,\textbf{u}_{j_{2}-1},
       \textbf{u}_{j_{2}+1},\ldots,\textbf{u}_s),\\ \nonumber\\
  \label{vdhi}
   &&\hskip-5mm b_{s}(t,\textbf{u}_1,\ldots,\textbf{u}_s)|_{t=0}=b_{s}^0(\textbf{u}_1,\ldots,\textbf{u}_s),
       \quad s\geq1,
\end{eqnarray}
where in recurrence evolution equations (\ref{vdh}) the operators $\Lambda^{[1]}(j)$ and
$\Lambda^{[2]}(j_1,j_{2})$ are determined by formula (\ref{gen_obs_gener}).

Further we consider initial states specified by a one-particle
marginal distribution function in the presence of correlations, namely
\begin{eqnarray}\label{lins}
   &&\hskip-8mm f^{(c)}\equiv(1,f_1^0(\textbf{u}_1),g_{2}(\textbf{u}_1,\textbf{u}_2)
        \prod_{i=1}^{2}f_{1}^0(\textbf{u}_i),\ldots,
        g_{s}(\textbf{u}_1,\ldots,\textbf{u}_s)\prod_{i=1}^{s}f_{1}^0(\textbf{u}_i),\ldots),
\end{eqnarray}
where the bounded functions $g_{s}\equiv g_{s}(\textbf{u}_1,\ldots,\textbf{u}_s),\,s\geq2$, are
specified initial correlations. Such assumption about initial states is intrinsic for the
kinetic description of complex systems in condensed states.

If $b(t)\in C_{\gamma}$ and $f_1^0\in L^{1}(\mathcal{J}\times\mathcal{U})$,
then under the condition that $\|f_1^0\|_{L^{1}(\mathcal{J}\times\mathcal{U})}<\gamma$,
there exists a mean field scaling limit of the mean value functional of marginal
observables and it is determined by the following series expansion:
\begin{eqnarray*}
   &&\hskip-7mm\big(b(t),f^{(c)}\big)=\sum\limits_{s=0}^{\infty}\,\frac{1}{s!}\,
      \int_{(\mathcal{J}\times\mathcal{U})^s}d\textbf{u}_{1}\ldots d\textbf{u}_{s}
      \,b_{s}(t,\textbf{u}_1,\ldots,\textbf{u}_s)g_{s}(\textbf{u}_1,\ldots,\textbf{u}_s)
      \prod\limits_{i=1}^{s} f_1^0(\textbf{u}_i).
\end{eqnarray*}

Then for the mean value functionals of the limit initial additive-type marginal observables,
i.e. of the sequences $b^{(1)}(0)=(0,b_{1}^0(\textbf{u}_1),0,\ldots)$~\cite{BG}, the
following representation is true:
\begin{eqnarray}\label{avmar-2}
   &&\hskip-7mm\big(b^{(1)}(t),f^{(c)}\big)=\sum\limits_{s=0}^{\infty}\,\frac{1}{s!}\,
       \int_{(\mathcal{J}\times\mathcal{U})^s}d\textbf{u}_{1}\ldots d\textbf{u}_{s}
       \,b_{s}^{(1)}(t,\textbf{u}_1,\ldots,\textbf{u}_s)g_{s}(\textbf{u}_1,\ldots,\textbf{u}_s)
       \prod\limits_{i=1}^{s} f_{1}^0(\textbf{u}_i)=\\
   &&\hskip-7mm\int_{(\mathcal{J}\times\mathcal{U})}d\textbf{u}_{1}\,
       b_{1}^{0}(\textbf{u}_{1})f_{1}(t,\textbf{u}_{1}).\nonumber
\end{eqnarray}
In equality (\ref{avmar-2}) the function $b_{s}^{(1)}(t)$ is given by a special case of
expansion (\ref{Iterd}), namely
\begin{eqnarray*}\label{itvad}
   &&\hskip-5mm b_{s}^{(1)}(t,\textbf{u}_1,\ldots,\textbf{u}_s)=\\
   &&\hskip-2mm\int_0^t dt_{1}\ldots\int_0^{t_{s-2}}dt_{s-1}
       \,e^{(t-t_{1})\sum\limits_{k_{1}=1}^{s}\Lambda^{[1]}(k_{1})}
       \sum\limits_{i_{1}\neq j_{1}=1}^{s}\Lambda^{[2]}(i_{1},j_{1})\,
       e^{(t_{1}-t_{2})\sum\limits_{l_{1}=1,\,\,l_{1}\neq j_{1}}^{s}\Lambda^{[1]}(l_{1})}\\
   &&\ldots\,e^{(t_{s-2}-t_{s-1})\sum\limits_{k_{s-1}=1,\,\,k_{s-1}\neq (j_{1},\ldots,j_{s-2})}^{s}
       \Lambda^{[1]}(k_{s-1})}\sum\limits^{s}_{\mbox{\scriptsize $\begin{array}{c}i_{s-1}\neq j_{s-1}=1,\\
       i_{s-1},j_{s-1}\neq (j_{1},\ldots,j_{s-2})\end{array}$}}\Lambda^{[2]}(i_{s-1},j_{s-1})\\
   &&\times e^{t_{s-1}\sum\limits_{l_{s-1}=1,\,\,l_{s-1}\neq (j_{1},\ldots,j_{s-1})}^{s}
       \Lambda^{[1]}(l_{s-1})}\,
       b_{1}^{0}((\textbf{u}_1,\ldots,\textbf{u}_s)\setminus(\textbf{u}_{j_{1}},\ldots,\textbf{u}_{j_{s-1}})),
       \quad s\geq1,
\end{eqnarray*}
and the limit one-particle distribution function $f_{1}(t)$ is represented
by the series expansion
\begin{eqnarray}\label{viter}
   &&\hskip-9mm f_{1}(t,\textbf{u}_1)=\sum\limits_{n=0}^{\infty}\int_0^tdt_{1}\ldots\int_0^{t_{n-1}}dt_{n}
      \int_{(\mathcal{J}\times\mathcal{U})^n}d \textbf{u}_{2}\ldots d \textbf{u}_{n+1}\,
      e^{(t-t_{1})\Lambda^{\ast[1]}(1)}\times\\
   &&\times\Lambda^{\ast[2]}(1,2)
      \prod\limits_{j_1=1}^{2}e^{(t_{1}-t_{2})\Lambda^{\ast[1]}(j_1)}\ldots
      \prod\limits_{j_{n-1}=1}^{n}e^{(t_{n-1}-t_{n})\Lambda^{\ast[1]}(j_{n-1})}\times\nonumber\\
   &&\times\sum\limits_{i_{n}=1}^{n}\Lambda^{\ast[2]}(i_{n},n+1)
      \prod\limits_{j_n=1}^{n+1}e^{t_{n}\Lambda^{\ast[1]}(j_{n})}
      g_{1+n}(\textbf{u}_1,\ldots,\textbf{u}_{n+1})\prod\limits_{i=1}^{n+1}f_{1}^0(\textbf{u}_i),\nonumber
\end{eqnarray}
where the operators $\Lambda^{\ast[i]},\,i=1,2,$ are adjoint operators (\ref{gen_state_gener})
to the operators $\Lambda^{[i]},\,i=1,2$ defined by formula (\ref{gen_obs_gener}), and on the
space $L^{1}_{n}$ these operators are defined as follows:
\begin{eqnarray*}
 &&\hskip-8mm \Lambda^{\ast[1]}(i) f_n(\textbf{u}_1,\ldots,\textbf{u}_n)\doteq
     \int_{\mathcal{J}\times\mathcal{U}}
     A^{[1]}(\textbf{u}_{i};\textbf{v})a^{[1]}(\textbf{v})\times\\
  &&\times f_n(\textbf{u}_1,\ldots,
     \textbf{u}_{{i}-1},\textbf{v},\textbf{u}_{{i}+1},\ldots,\textbf{u}_n)d\textbf{v}-
     a^{[1]}(\textbf{u}_{i})f_n(\textbf{u}_1,\ldots,\textbf{u}_n)\nonumber,\\
 &&\hskip-8mm \Lambda^{\ast[2]}(i,j)f_n(\textbf{u}_1,\ldots,\textbf{u}_n)\doteq
    \int_{\mathcal{J}\times\mathcal{U}}
     A^{[2]}(\textbf{u}_{i};\textbf{v},\textbf{u}_{j})
     a^{[2]}(\textbf{v},\textbf{u}_{j})\times\nonumber\\
 &&\times f_n(\textbf{u}_1,\ldots,\textbf{u}_{{i}-1},\textbf{v},
     \textbf{u}_{{i}+1},\ldots,\textbf{u}_n)d\textbf{v}-
     a^{[2]}(\textbf{u}_{i},\textbf{u}_{j})f_n(\textbf{u}_1,\ldots,\textbf{u}_n)\nonumber,
\end{eqnarray*}
where the functions $A^{[m]},a^{[m]},\,m=1,2$, are defined above in formula (\ref{gen_obs_gener}).

For initial data $f_{1}^0\in L^{1}(\mathcal{J}\times\mathcal{U})$ limit marginal distribution
function (\ref{viter}) is a strong solution of the Cauchy problem of the Vlasov-type kinetic equation
with initial correlations:
\begin{eqnarray}\label{Vlasov1}
    &&\hskip-7mm\frac{\partial}{\partial t}f_{1}(t,\textbf{u}_1)= \Lambda^{\ast[1]}(1)f_{1}(t,\textbf{u}_1)+\\
    &&+\int_{\mathcal{J}\times\mathcal{U}}d\textbf{u}_{2}
       \Lambda^{\ast[2]}(1,2)\prod_{i_1=1}^{2}e^{t\Lambda^{\ast[1]}(i_1)}g_{2}(\textbf{u}_1,\textbf{u}_2)
     \prod_{i_2=1}^{2}e^{-t\Lambda^{\ast[1]}(i_2)}
     f_{1}(t,\textbf{u}_1)f_{1}(t,\textbf{u}_2),\nonumber\\ \nonumber\\
  \label{Vlasov2}
    &&\hskip-7mmf_{1}(t,\textbf{u}_1)|_{t=0}=f_1^0(\textbf{u}_1),
\end{eqnarray}
where the function $g_{2}(\textbf{u}_1,\textbf{u}_2)$ is initial correlation function specified initial
state (\ref{lins}).

For mean value functionals of the limit initial $k$-ary marginal observables, i.e. of
the sequences $b^{(k)}(0)=(0,\ldots,0,b_{k}^0(\textbf{u}_1,\ldots,\textbf{u}_k),0,\ldots)$,
the following representation is true:
\begin{eqnarray}\label{dchaos}
    &&\hskip-8mm\big(b^{(k)}(t),f^{(c)}\big)=\sum\limits_{s=0}^{\infty}\,\frac{1}{s!}\,
       \int_{(\mathcal{J}\times\mathcal{U})^s}d\textbf{u}_{1}\ldots d\textbf{u}_{s}
       \,b_{s}^{(k)}(t,\textbf{u}_1,\ldots,\textbf{u}_s) g_{s}(\textbf{u}_1,\ldots,\textbf{u}_s)
       \prod\limits_{i=1}^{s} f_1^0(\textbf{u}_i)=\\
    &&\hskip-5mm=\frac{1}{k!}\int_{(\mathcal{J}\times\mathcal{U})^k}d\textbf{u}_{1}\ldots d\textbf{u}_{k}
       \,b_{k}^0(\textbf{u}_1,\ldots,\textbf{u}_k)\times\nonumber\\
     &&\hskip-5mm\times\prod_{i_1=1}^{k}e^{t\Lambda^{\ast[1]}(i_1)}
       g_{k}(\textbf{u}_1,\ldots,\textbf{u}_k)\prod_{i_2=1}^{k}e^{-t\Lambda^{\ast[1]}(i_2)}
       \prod\limits_{i=1}^{k} f_{1}(t,\textbf{u}_i),\quad k\geq2,\nonumber
\end{eqnarray}
where the limit one-particle marginal distribution function $f_{1}(t,\textbf{u}_i)$ is determined
by series expansion (\ref{viter}) and the functions $g_{k}(\textbf{u}_1,\ldots,\textbf{u}_k),\,k\geq2,$
are initial correlation functions specified initial state (\ref{lins}).

Hence in case of $k$-ary marginal observables the evolution governed by the dual Vlasov
hierarchy (\ref{vdh}) is equivalent to a property of the propagation of initial correlations
(\ref{dchaos}) for the $k$-particle marginal distribution function or in other words mean
field scaling dynamics does not create correlations.

In case of initial states of statistically independent entities specified by a one-particle
marginal distribution function, namely $f^{(c)}\equiv(1,f_1^0(\textbf{u}_1),\ldots,
{\prod\limits}_{i=1}^{s}f_{1}^0(\textbf{u}_i),\ldots)$, the kinetic evolution of $k$-ary
marginal observables governed by the dual Vlasov hierarchy means the property of the propagation
of initial chaos for the $k$-particle marginal distribution function within the framework of the
evolution of states \cite{CIP}, i.e. a sequence of the limit distribution functions has the form $f(t)\equiv(1,f_1(t,\textbf{u}_1),\ldots,{\prod\limits}_{i=1}^{s}f_{1}(t,\textbf{u}_i),\ldots)$,
where the one-particle distribution function $f_1(t)$ is governed by the Vlasov kinetic equation \cite{GF13}
\begin{eqnarray*}
      &&\hskip-5mm\frac{\partial}{\partial t}f_{1}(t,\textbf{u}_1)= \Lambda^{\ast[1]}(1)f_{1}(t,\textbf{u}_1)+
       \int_{\mathcal{J}\times\mathcal{U}}d\textbf{u}_{2}
       \Lambda^{\ast[2]}(1,2)f_{1}(t,\textbf{u}_1)f_{1}(t,\textbf{u}_2).
\end{eqnarray*}

We note that, according to equality (\ref{dchaos}), in the mean field limit the marginal
correlation functions defined as cluster expansions of marginal distribution functions
\cite{G12},\cite{GP}, namely,
\begin{eqnarray*}
   &&f_{s}(t,\textbf{u}_1,\ldots,\textbf{u}_s)=
      \sum\limits_{\mbox{\scriptsize$\begin{array}{c}\mathrm{P}:(\textbf{u}_1,\ldots,\textbf{u}_s)=\bigcup_{i}\textbf{U}_{i}\end{array}$}}
      {\prod\limits}_{\textbf{U}_i\subset\mathrm{P}}g_{|\textbf{U}_i|}(t,\textbf{U}_i),\quad s\geq1,
\end{eqnarray*}
has the following explicit form \cite{GF15}:
\begin{eqnarray}\label{cfmf}
    &&\hskip-5mm g_{1}(t,\textbf{u}_1)=f_{1}(t,\textbf{u}_1),\\
    &&\hskip-5mm g_{s}(t,\textbf{u}_1,\ldots,\textbf{u}_s)=\prod_{i_1=1}^{s}e^{t\Lambda^{\ast[1]}(i_1)}
        \tilde{g}_{s}(\textbf{u}_1,\ldots,\textbf{u}_s)\prod_{i_2=1}^{s}e^{-t\Lambda^{\ast[1]}(i_2)}
        \prod\limits_{j=1}^{s}f_{1}(t,\textbf{u}_j),\quad s\geq2,\nonumber
\end{eqnarray}
where for initial correlation functions (\ref{lins}) it is used the following notations:
\begin{eqnarray*}\label{FG1}
   &&\hskip-5mm \tilde{g}_{s}(\textbf{u}_1,\ldots,\textbf{u}_s)=
      \sum\limits_{\mbox{\scriptsize$\begin{array}{c}\mathrm{P}:(\textbf{u}_1,\ldots,\textbf{u}_s)=\bigcup_{i}\textbf{U}_{i}\end{array}$}}
      \prod_{\textbf{U}_i\subset \mathrm{P}}g_{|\textbf{U}_i|}(\textbf{U}_i),
\end{eqnarray*}
the symbol $\sum_\mathrm{P}$ means the sum over possible partitions $\mathrm{P}$ of the set of arguments
$(\textbf{u}_1,\ldots,\textbf{u}_s)$ on $|\mathrm{P}|$ nonempty  subsets $\textbf{U}_i$, and
the one-particle marginal distribution function $f_{1}(t)$ is a solution of the Cauchy problem
of the Vlasov-type kinetic equation with initial correlations (\ref{Vlasov1}),(\ref{Vlasov2}).

Thus, an equivalent approach to the description of the kinetic evolution of large number
of interacting constituents in terms of the Vlasov-type kinetic equation with correlations
(\ref{Vlasov1}) is given by the dual Vlasov hierarchy (\ref{vdh}) for the additive-type
marginal observables.

\textcolor{blue!50!black}{\section{The non-Markovian generalized kinetic equation with initial correlations}}
Furthermore, the relationships between the evolution of observables of a large number
of interacting constituents of active soft condensed matter and the kinetic evolution
of its states described in terms of a one-particle marginal distribution function are discussed.

Since many-particle systems in condensed states are characterized
by correlations we consider initial states specified by a one-particle
marginal distribution function and correlation functions, namely
\begin{eqnarray}\label{ch}
   &&F^{(c)}=(1,F_{1}^{0,\varepsilon}(\textbf{u}_1),g_{2}^{\varepsilon}(\textbf{u}_1,\textbf{u}_2)
      \prod_{i=1}^{2}F_{1}^{0,\varepsilon}(\textbf{u}_i),\ldots,g_{s}^{\varepsilon}(\textbf{u}_1,\ldots,\textbf{u}_s)
      \prod_{i=1}^{s}F_{1}^{0,\varepsilon}(\textbf{u}_i),\ldots).
\end{eqnarray}

If the initial state is completely specified by a one-particle distribution function and
a sequence of correlation functions (\ref{ch}), then, using a nonperturbative solution of
the dual BBGKY hierarchy (\ref{sdh}), in \cite{GF14} it was proved that all possible states
at the arbitrary moment of time can be described within the framework of a one-particle distribution
function governed by the non-Markovian generalized kinetic equation with initial correlations,
i.e. without any approximations like in scaling limits as above.

Indeed, for initial states (\ref{ch}) for mean value functional (\ref{avmar}) the equality holds
\begin{eqnarray}\label{w}
    &&\big(B(t),F^{(c)}\big)=\big(B(0),F(t\mid F_{1}(t))\big),
\end{eqnarray}
where $F(t\mid F_{1}(t))=(1,F_1(t),F_2(t\mid F_{1}(t)),\ldots,F_s(t\mid F_{1}(t)),\ldots)$ is
a sequence of marginal functionals of the state with respect to a one-particle
marginal distribution function
\begin{eqnarray}\label{F_1(t)}
    &&\hskip-12mm F_{1}(t,\textbf{u}_1)=\sum\limits_{n=0}^{\infty}\frac{1}{n!}
        \int\limits_{(\mathcal{J}\times\mathcal{U})^n}d\textbf{u}_{2}\ldots d\textbf{u}_{n+1}
        \mathfrak{A}_{1+n}^\ast(t,1,\ldots,n+1)g_{n+1}^{\varepsilon}(\textbf{u}_1,\ldots,\textbf{u}_{n+1})
        \prod_{i=1}^{n+1}F_{1}^{0,\varepsilon}(\textbf{u}_i).
\end{eqnarray}
The generating operator $\mathfrak{A}_{1+n}^\ast(t)$ of series (\ref{F_1(t)}) is the $(1+n)$-order cumulant of
the semigroups of operators $\{e^{t\Lambda^\ast_{n}}\}_{t\geq0},\,n\geq1$.

The marginal functionals of the state is defined by the series expansions:
\begin{eqnarray}\label{f}
   &&\hskip-12mm F_{s}\big(t,\textbf{u}_1,\ldots,\textbf{u}_s \mid F_{1}(t)\big)\doteq
      \sum_{n=0}^{\infty }\frac{1}{n!}\int\limits_{(\mathcal{J}\times\mathcal{U})^n}d\textbf{u}_{s+1}\ldots
      d\textbf{u}_{s+n}\,\mathfrak{V}_{1+n}(t,\{Y\},X\setminus Y)\prod_{i=1}^{s+n}F_{1}(t,\textbf{u}_i),
\end{eqnarray}
where the following notations used: $Y\equiv(1,\ldots,s)$,\,\,$X\setminus Y\equiv(s+1,\ldots,s+n)$ and
the generating operators $\mathfrak{V}_{1+n}(t),\,n\geq0$, are defined by the expansions \cite{GF14}:
\begin{eqnarray}\label{skrrn}
    &&\hskip-8mm\mathfrak{V}_{1+n}(t,\{Y\},X\setminus Y)\doteq\sum_{k=0}^{n}(-1)^k\,\sum_{m_1=1}^{n}\ldots
       \sum_{m_k=1}^{n-m_1-\ldots-m_{k-1}}\frac{n!}{(n-m_1-\ldots-m_k)!}\times\\
    &&\hskip-5mm \times\widehat{\mathfrak{A}}_{1+n-m_1-\ldots-m_k}(t,\{Y\},s+1,
       \ldots,s+n-m_1-\ldots-m_k)\prod_{j=1}^k\,\sum_{k_2^j=0}^{m_j}\ldots\nonumber\\
    &&\hskip-5mm\sum_{k^j_{n-m_1-\ldots-m_j+s}=0}^{k^j_{n-m_1-\ldots-m_j+s-1}}\,\prod_{i_j=1}^{s+n-m_1-\ldots-m_j}
       \frac{1}{(k^j_{n-m_1-\ldots-m_j+s+1-i_j}-k^j_{n-m_1-\ldots-m_j+s+2-i_j})!}\nonumber\\
    &&\hskip-5mm \times\widehat{\mathfrak{A}}_{1+k^j_{n-m_1-\ldots-m_j+s+1-i_j}-k^j_{n-m_1-\ldots-m_j+s+2-i_j}}(t,
       i_{j},s+n-m_1-\ldots-m_j+1\nonumber \\
    &&\hskip-5mm+k^j_{s+n-m_1-\ldots-m_j+2-i_j},\ldots,s+n-m_1-\ldots-m_j+k^j_{s+n-m_1-\ldots-m_j+1-i_j}),\nonumber
\end{eqnarray}
where $k^j_1\equiv m_j,$ і $k^j_{n-m_1-\ldots-m_j+s+1}\equiv 0$ and the evolution operators
$\widehat{\mathfrak{A}}_{n}(t),\,n\geq1,$ are cumulants of the semigroups of scattering
operators $\{e^{t\Lambda_{k}^{\ast}}g_{k}^{\varepsilon}\prod_{i=1}^{k}e^{-t\Lambda^{\ast[1]}(i)}\}_{t\geq0},\,k\geq1$.
We adduce some examples of evolution operators (\ref{skrrn}):
\begin{eqnarray*}\label{rrrls}
   &&\hskip-12mm\mathfrak{V}_{1}(t,\{Y\})=\widehat{\mathfrak{A}}_{1}(t,\{Y\})\doteq
       e^{t\Lambda^{\ast}_s}g_{s}^{\varepsilon}\prod_{i=1}^{s}e^{-t\Lambda^{\ast[1]}(i)},\\
   &&\hskip-12mm\mathfrak{V}_{2}(t,\{Y\},s+1)=\widehat{\mathfrak{A}}_{2}(t,\{Y\},s+1)-
       \widehat{\mathfrak{A}}_{1}(t,\{Y\})\sum_{i_1=1}^s\widehat{\mathfrak{A}}_{2}(t,i_1,s+1).
\end{eqnarray*}

If $\|F_{1}(t)\|_{L^{1}(\mathcal{J}\times\mathcal{U})}<e^{-(3s+2)}$, then for arbitrary $t\in \mathbb{R}$
series expansion (\ref{f}) converges in the norm of the space $L^{1}_{s}$ \cite{G12}.

The proof of equality (\ref{w}) is based on the application of cluster expansions
to generating operators (\ref{cumulantd}) of expansions (\ref{sdh}) which are dual to the
kinetic cluster expansions introduced in paper \cite{GG}. Then the adjoint series expansion
can be expressed in terms of one-particle distribution function (\ref{F_1(t)}) in the
form of the functional from the right-hand side of equality (\ref{w}).

We emphasize that marginal functionals of the state (\ref{f}) characterize the processes of the creation
of correlations generated by dynamics of many-constituent systems of active soft condensed matter and
the propagation of initial correlations.

For small initial data $F_1^{0,\epsilon}\in L^{1}(\mathcal{J}\times\mathcal{U})$~\cite{GF14},
series expansion (\ref{F_1(t)}) is a global in time solution of the Cauchy problem of the generalized
kinetic equation with initial correlations:
\begin{eqnarray}
  \label{gke}
    &&\hskip-8mm\frac{\partial}{\partial t}F_{1}(t,\textbf{u}_1)= \Lambda^{\ast[1]}(1)F_{1}(t,\textbf{u}_1)+\\
    &&\hskip-8mm+\sum\limits_{k=1}^{M-1}\frac{\varepsilon^{k}}{k!}
       \int\limits_{(\mathcal{J}\times\mathcal{U})^k}d\textbf{u}_{2}\ldots d\textbf{u}_{k+1}
       \hskip-2mm\sum\limits_{\mbox{\scriptsize $\begin{array}{c}j_1\neq\ldots\neq j_{k+1}\in\\
       \in(1,\ldots,k+1)\end{array}$}}\hskip-2mm\Lambda^{\ast[k+1]}(j_1,\ldots,j_{k+1})
       F_{k+1}\big(t,\textbf{u}_1,\ldots,\textbf{u}_{k+1}\mid F_{1}(t)\big),\nonumber\\ \nonumber\\
  \label{gkei}
    &&\hskip-8mm F_{1}(t,\textbf{u}_1)|_{t=0}=F_1^{0,\epsilon}(\textbf{u}_1).
\end{eqnarray}
For initial data $F_1^{0,\epsilon}\in L^{1}(\mathcal{J}\times\mathcal{U})$ it is a strong (classical) solution
and for an arbitrary initial data it is a weak (generalized) solution.

In particular case $M=2$ of two subpopulations kinetic equation (\ref{gke}) has the following explicit form:
\begin{eqnarray*}
    &&\hskip-8mm\frac{\partial}{\partial t}F_{1}(t,\textbf{u}_1)=
       \int\limits_{\mathcal{J}\times\mathcal{U}} A^{[1]}(\textbf{u}_{1};\textbf{v})
       a^{[1]}(\textbf{v})F_1(t,\textbf{v})d\textbf{v}-a^{[1]}(\textbf{u}_{1})F_1(t,\textbf{u}_{1})+ \\
    &&\hskip-8mm\int\limits_{\mathcal{J}\times\mathcal{U}}d\textbf{u}_{2}\Big(
       \int\limits_{\mathcal{J}\times\mathcal{U}}A^{[2]}(\textbf{u}_{1};\textbf{v},\textbf{u}_{2})
        a^{[2]}(\textbf{v},\textbf{u}_{2})
        F_{2}\big(t,\textbf{v},\textbf{u}_{2}\mid F_{1}(t)\big)d\textbf{v}-a^{[2]}(\textbf{u}_{1},\textbf{u}_{2})
        F_{2}\big(t,\textbf{u}_{1},\textbf{u}_{2}\mid F_{1}(t)\big)\Big),
\end{eqnarray*}
where the functions $A^{[k]}$ and $a^{[k]}$ are defined above.

We note that for initial states (\ref{ch}) specified by a one-particle (marginal) distribution function,
the evolution of states described within the framework of a one-particle (marginal)
distribution function governed by the generalized kinetic equation with initial correlations
(\ref{gke}) is dual to the dual BBGKY hierarchy for additive-type marginal observables
with respect to bilinear form (\ref{averageD}), and it is completely equivalent to the description
of states in terms of marginal distribution functions governed by the BBGKY hierarchy of interacting
entities.

Thus, the evolution of many-constituent systems of active soft condensed matter described in terms
of marginal observables in case of initial states (\ref{ch}) can be also described within the framework
of a one-particle (marginal) distribution function governed by the non-Markovian generalized kinetic
equation with initial correlations (\ref{gke}).

We remark, considering that a mean field limit of initial state (\ref{ch}) is described by sequence
(\ref{lins}), a mean field asymptotics of a solution of the non-Markovian generalized kinetic equation
with initial correlations (\ref{gke}) is governed by the Vlasov-type kinetic equation with initial
correlations (\ref{Vlasov1}) derived above from the dual Vlasov hierarchy (\ref{vdh}) for limit marginal
observables of interacting entities \cite{GF15}. Moreover, a mean field asymptotic behavior of marginal
functionals of the state (\ref{f}) describes the propagation in time of initial correlations like
established property (\ref{cfmf}).

\textcolor{blue!50!black}{\section{Conclusion}}

We considered an approach to the description of kinetic evolution of large number
of interacting constituents (entities) of active soft condensed matter within the framework
of the evolution of marginal observables of these systems. Such representation of the
kinetic evolution seems, in fact, the direct mathematically fully consistent formulation
modeling the collective behavior of biological systems since the notion of state is more
subtle and implicit characteristic of living creatures.

A mean field scaling asymptotics of nonperturbative solution (\ref{sdh}) of the dual
BBGKY hierarchy (\ref{dh}) for marginal observables was constructed. The constructed
scaling limit of a nonperturbative solution (\ref{sdh}) is governed by the set of
recurrence evolution equations (\ref{vdh}), namely, by the dual Vlasov hierarchy for
interacting stochastic processes modeling large particle systems of active soft condensed
matter.

We established that the limit additive-type marginal observables governed by the dual
Vlasov hierarchy (\ref{vdh}) gives an equivalent approach to the description of the
kinetic evolution of many entities in terms of a one-particle distribution function governed
by the Vlasov kinetic equation with initial correlations (\ref{Vlasov1}). Moreover, the kinetic
evolution of non-additive type marginal observables governed by the dual Vlasov hierarchy
means the property of the propagation of initial correlations (\ref{cfmf}) within the framework
of the evolution of states.

One of the advantages of suggested approach in comparison with the conventional approach
of the kinetic theory \cite{CGP97},\cite{V},\cite{CIP} is the possibility to construct kinetic
equations in various scaling limits in the presence of initial correlations which can characterize
the analogs of condensed states of many-particle systems of statistical mechanics for interacting
entities of complex biological systems.

We note that the developed approach is also related to the problem of a rigorous derivation
of the non-Markovian kinetic-type equations from underlying many-entity dynamics which
make it possible to describe the memory effects of collective dynamics of complex systems
modeling active soft condensed matter.

In case of initial states completely specified by a one-particle distribution function and
correlations (\ref{ch}), using a nonperturbative solution of the dual BBGKY hierarchy (\ref{sdh}),
it was proved that all possible states at the arbitrary moment of time can be described within the
framework of a one-particle distribution function governed by the non-Markovian generalized kinetic
equation with initial correlations (\ref{gke}), i.e. without any approximations. A mean field asymptotics
of a solution of kinetic equation with initial correlations (\ref{gke}) is governed by the Vlasov-type
kinetic equation with initial correlations (\ref{Vlasov1}) derived above from the dual Vlasov hierarchy
(\ref{vdh}) for limit marginal observables.

Moreover, in the case under consideration the processes of the creation of correlations generated by dynamics
of large particle systems of active soft condensed matter and the propagation of initial correlations are
described by the constructed marginal functionals of the state (\ref{f}) governed by the non-Markovian generalized
kinetic equation with initial correlations (\ref{gke}).


\vspace{0.5cm}
\addcontentsline{toc}{section}{{References}}
\renewcommand{\refname}{\textcolor{blue!50!black}{References}}
\small{

}
\end{document}